\documentclass[aps,prl,twocolumn,amsmath,amssymb,superscriptaddress,nobibnotes,showpacs,floatfix]{revtex4-1}
\usepackage{graphicx,SIunits,array,multirow,amsmath,amsfonts,tabularx}
\usepackage{dsfont,color}

\newcolumntype{Y}{>{\centering\arraybackslash}X}

\begin{document}

\title{A new route to relativistic insulators: Lifshitz transition driven by spin fluctuations and spin-orbit renormalization in NaOsO$_3$}

\author{Bongjae Kim}
\affiliation{University of Vienna, Faculty of Physics and Center for Computational Materials Science, Vienna, Austria}

\author{Peitao Liu}
\affiliation{University of Vienna, Faculty of Physics and Center for Computational Materials Science, Vienna, Austria}

\author{Zeynep Erg\"{o}nenc}
\affiliation{University of Vienna, Faculty of Physics and Center for Computational Materials Science, Vienna, Austria}

\author{Alessandro Toschi}
\affiliation{Institut f\"{u}r Festk\"{o}rperphysik, Technische Universit\"{a}t Wien, Vienna, Austria}

\author{Sergii Khmelevskyi}
\affiliation{University of Vienna, Faculty of Physics and Center for Computational Materials Science, Vienna, Austria}
\affiliation{Institute for Applied Physics, Technische Universit\"{a}t Wien, Vienna, Austria}

\author{Cesare Franchini}\email{cesare.franchini@univie.ac.at}
\affiliation{University of Vienna, Faculty of Physics and Center for Computational Materials Science, Vienna, Austria}

\date{\today}

\begin{abstract}
In systems where electrons form both dispersive bands and small local spins, we show that changes of the
spin configuration can tune the bands through a Lifshitz transition, resulting in a continuous
metal-insulator transition associated with a progressive change of the Fermi surface topology.
In contrast to a Mott-Hubbard and Slater pictures,
this spin-driven Lifshitz transition appears in systems with small electron-electron correlation and
large hybridization. We show that this situation is realized in 5$d$ distorted perovskites with an half-filled $t_{2g}$ bands such as NaOsO$_3$,
where the strong $p-d$ hybridization reduces the local moment, and spin-orbit coupling
causes a large renormalization of the electronic mobility. This weakens the role of electronic
correlations and drives the system towards an itinerant magnetic regime which enables spin-fluctuations.
\end{abstract}

\pacs{71.30.+h, 75.47.Lx, 71.27.+a}

\maketitle

 Metal-to-insulator transitions (MITs) are one of the most important phenomena in solid-state physics and their
 fundamental understanding represents an enduring challenge in solid state theory~\cite{Mott1968,Imada1998}.
 Different mechanisms have been invoked to explain the formation of an insulating regime.
 Classical examples are realized in 3$d$ transition metal oxides (TMOs), where the nonconducting state is typically understood
 within the Mott-Hubbard model as arising from the competition between
 strong electron-electron interaction ($U$) and the electronic
 mobility, associated to the (non-interacting) bandwidth ($W$)~\cite{Dagotto2005,Tokura2000,Kotliar2004}.
 When moving to the more spatially extended 4$d$ and 5$d$ orbitals the $W$ increases and the $U$ is expected to become smaller,
 leading to the tendency towards metallicity as in the itinerant magnet SrRuO$_3$~\cite{Koster2012}.
 In contrast to these expectations, recent theory and experiment have revealed that in 'heavy' 5$d$ TMOs,
 the enhanced spin-orbit coupling (SOC) strength~\cite{Cao1998} can lead to the formation of a variety of novel types of quantum states
 including unexpected insulating regimes~\cite{BJKim2008,Jackeli2009,Pesin2010,YKKim2015}.
 In the most representative example, Sr$_2$IrO$_4$, the concerted action of a strong SOC and a moderate $U$ leads to the opening
 of a small spectral gap~\cite{BJKim2008, Martins2011} denominated {\sl relativistic Mott} gap.
 Other and more rare types of MITs have been recently proposed for magnetic relativistic osmium oxides based on the
 Slater mechanism~\cite{Shi2009,Calder2012,LoVecchio2013,Calder2015}, driven by antiferromagnetic (AFM) order, where the gap is opened
 by exchange interactions and not by electronic correlation,~\cite{Slater1951} or Lifshitz-like processes~\cite{Hiroi2015, Sohn2015},
 involving a rigid change of the Fermi surface topology~\cite{Lifshitz1960}.

 In this work we show that in weakly correlated (small $U$) itinerant magnets the combined effect of longitudinal and rotational spin-fluctuations
 can cause a continuous MIT with Lifshitz characteristics, fundamentally different from relativistic Mott or purely Slater insulating states.
 The necessary conditions for the onset of this type of spin-driven Lifshitz MIT are the coexistence of a small $U$, a small local moment and an high
 degree of orbital hybridization. This places the system at the border of a magnetic and electronic instability where a Lifshitz MIT is possible.
 This situation can be realized in structurally-distorted, half-filled $t_{2g}^3$ (or close to half-filled) 5$d$ oxides such as NaOsO$_3$
 and Cd$_2$Os$_2$O$_7$ or iridates.
 By taking NaOsO$_3$ as a prototypical example, we show that here the balance between
 $U$ and $W$ is controlled by the SOC: the SOC induces a surprisignly large reduction of electronic mobility which renormalizes the $U$ and pushes the system
 into a weakly correlated and magnetically itinerant regime subjected to spin fluctuations.
 We find that at high-temperature NaOsO$_3$ is a paramagnetic metal
 but as temperature decreases the changes of the amplitue and direction of the spins lead to the continuous vanishing of holes and
 electrons pockets in the Fermi surface (FS), that do not involve any substantial modification of the underlying band structure~\cite{Lifshitz1960},
 consistent with a Lifshitz-type MIT.

 Experimental observations indicate that NaOsO$_3$ undergoes a second-order MIT concomitant with the onset of an
 antiferromagnetic long-range ordering at a N\'{e}el temperature ($T_N$) of $\approx$ 410~K. This behavior is
 apparently adaptable to a Slater mechanism~\cite{Shi2009,Calder2012,LoVecchio2013, Calder2015, Jung2013}.
 However, the bad-metal behavior observed in a wide intermediate temperature region ($0.1 T_N<T<T_N$)~\cite{Shi2009},
 the Curie-Weiss behavior of the susceptibility above $T_N$, and the need of a sizable $U$ in Density Functional Theory (DFT) calculations to
 open the gap, is in conflict with an authentic Slater mechanism~\cite{Du2012,Jung2013,Middey2014}.

 The fundamental properties of NaOsO$_3$ have been clearly exposed by Jung {\em {et al.}}~\cite{Jung2013}.
 In particular, it was shown that the apparent discrepancy between the measured Os ordered moment of only
 1 $\mu_B$\cite{Calder2012} and the nominal $t_{2g}^3$ configuration should be attributable to the large $p-d$ hybridization
 that effectively reduces the local moment and form an (electronic and magnetic) itinerant background~\cite{Jung2013}.
 However, the authors concluded that the role of SOC is modest in this system, also based on the, reasonable, consideration that
 the $t_{2g}^{3}$ associated with a $L_{eff}=0$ state exhibits a negligible orbital moment,
 much smaller then the one of partially filled $t_{2g}$ subshells~\cite{Gangopadhyay,Liu2015}.
 In the following we will show that relativistic effects can be strong even in a quenched orbital moment scenario,
 and we will explain that the SOC is the crucial effect in paving the way for the continuous MIT in  NaOsO$_3$.

\emph{Spin-orbit induced renormalization.} In the presence of a MIT in a TMO, the clarification of the role
played by the electronic correlation $U$ is a central aspect for the
theoretical understanding. We begin, hence, by studying the effects of $U$ and its
interplay with the SOC for the case of our interest, NaOsO$_3$. All calculations were performed
using the Vienna \emph{ab initio} simulation package (VASP) \cite{Kresse1996} using the DFT+$U$ method
and including relativistic effects (see Supplement).

\begin{figure}[t]
\includegraphics[width=0.45\textwidth]{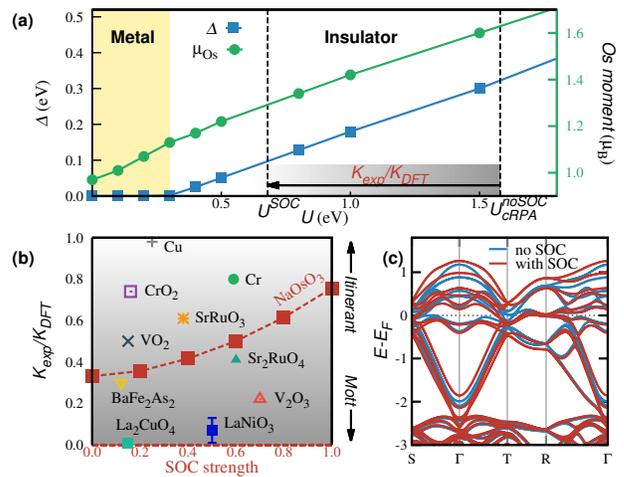}
\caption{(color online). {Relativistic ``renormalization'' of the electronic correlation.}
({a}) Energy gap and Os magnetic moment.
({b}) $K_{exp}/K_{DFT}$ as a function of the SOC strength for
NaOsO$_3$ and other reference systems.
The measured data of the electron mobility are taken from Ref. \protect\cite{LoVecchio2013} (NaOsO$_3$),
Ref. \protect\cite{Stewart2011} (LaNiO$_3$), and Ref.  \protect\cite{Qazilbash2009}
(all other materials).
({c}) Paramagnetic band structure of NaOsO$_3$ with and without SOC.
}
\label{Corr}
\end{figure}

We start by computing the dependence of experimentally accessible
observables on $U$.
As usual, the value of the spectral gap $\Delta$ and the magnetic moment $\mu _{Os}$
are highly sensitive on the choice of $U$, as visualized in Figure~\ref{Corr}(a).
For $U=0$ the system is metallic, but for $U\geq0.3$ eV NaOsO$_3$ exhibits a finite gap that increases linearly with $U$. The experimentally
reported low-temperature optical gap, 0.1~eV~\cite{LoVecchio2013} and the $\mu _{Os}$ $\approx$ 1 $\mu_B$~\cite{Shi2009} clearly indicate that the
optimal value of $U$ for NaOsO$_3$ should be chosen around $0.5$~eV.
At the same time, if we calculate $U$ entirely \emph{ab initio} within the constrained random phase approximation
(cRPA)~\cite{Aryasetiawan2004},  neglecting all relativistic effects
(SOC=0),  we obtain $U^{noSOC}=1.58$~eV, a value similar to the one used in previous studies
(i.e.,  $U=2$ eV~\cite{Du2012,Jung2013,Middey2014}).
As we demonstrate here, the root of this apparent controversy lays in
the suprising ``competition'' between SOC and electronic correlation.

We study the interplay between SOC and correlation
by comparing our DFT+$U$ results with available infrared optical experiments.
One rather general way\cite{Millis2005,Qazilbash2009,Si2009} to estimate the degree of electronic
correlation of a system is to evaluate  the reduction of the
electronic kinetic energy (mass renormalization) due to Coulomb repulsion.
This can be quantified by the ratio  $K_{exp}/K_{DFT}$  between the
experimentally measured kinetic energy  ($K_{exp}$), determined by
integrating over frequency the Drude contribution in the optical conductivity $\sigma _{D}$
and the correpsonding ``non-correlated'' kinetic energy $K_{DFT}$
obtained by DFT at $U=0$.
Considering that $\sigma _{D}$  can be expressed in terms of the
plasma frequency, $\omega _{p}$, via $\int d\omega \,
\sigma _{D}(\omega) =\omega _{p}^{2}/8$, one can rewrite the kinetic energy ratio as:
\begin{equation}
\frac{K_{exp}}{K_{DFT}}=\frac{\omega_{p,exp}^{2}}{\omega _{p,DFT}^{2}}
\end{equation}

Surprisingly, $K_{exp}/K_{DFT}$ in NaOsO$_{3}$ is dramatically dependent on the SOC strength as shown in Fig. \ref{Corr}(b).
Without SOC, one should have classified the system as
intermediate-strong correlated with $K_{exp}/K_{DFT}$ = 0.33, which is close to
Fe-pnictide superconductors. As SOC is considered in the DFT
calculations, however, $K_{DFT}$ gets systematically reduced, as the
estimated degree of correlation.  For a full inclusion of SOC (SOC=1) we obtain  $K_{exp}/K_{DFT}$ =0.76,
a value similar to those of conventional metals like Cr.
To our best knowledge, NaOsO$_{3}$ represents the first system where
such a strong renormalization of the balance between SOC and
correlations is reported.

Supporting evidence for the importance  of SOC in NaOsO$_3$ is provided by the computed SOC energy, 0.4 eV/Os, and by comparing the
electronic structure with and without SOC (Fig.~\ref{Corr}(b)). The inclusion of SOC leads to a widening of the
band width by about 0.3 eV, a linearization of the band near $\Gamma$ which yields the formation of a Dirac-like feature
and, most importantly for the electron mobility, a band flattening near $E_F$ that increases the effective masses and therefore
decreases $K_{DFT}$.

Therefore, by turning on the SOC  the estimated degree
of correlation moves gradually from the border to the Mott-physics to the one of conventional metallic systems.
This significant relativistic renormalization clarifies the origin of the abovementioned inconsistencies on the value of $U$ in this system.
More quantitatively, by rescaling the cRPA value of $U^{noSOC}$ by the SOC-induced renormalization factor~\cite{Si2009}
we obtain $U^{SOC}=0.68$ eV (comparable to the SOC energy) which leads to a much better description of the bandgap (see arrow in Fig. \ref{Corr}(a)).
This smaller value of $U$ is also consistent with the $U$ proposed for similar compounds like
LiOsO$_3$, $U<1$~eV~\cite{Giovannetti2014} and the itinerant magnet SrRuO$_3$, $U\approx 0.6$ eV~\cite{Rondinelli2008}.

Importantly, the analogy with SrRuO$_3$ is not limited to the low degree of correlation but also involves the magnetic itinerancy.
Although NaOsO$_3$ exhibits a high $T_N$=410~K, the effective moment extracted from the the Curie-Weiss behavior of the high temperature
susceptibility, 2.71~$\mu_B$~\cite{Shi2009}, is much higher than the ground state magnetic moment measured by neutron diffraction,
$\approx$ 1~$\mu_B$~\cite{Calder2012}, indicating a large Rhodes-Wohlfarth ratio and suggestive of an itinerant antiferromagnetic behavior~\cite{Mohn2006}.
This magnetic itinerancy would  {\sl not} be compatible with a large $U$ and is essential to explain the MIT in NaOsO$_3$, as discussed below.

\begin{figure}[t]
\includegraphics[angle=0,width=0.45\textwidth]{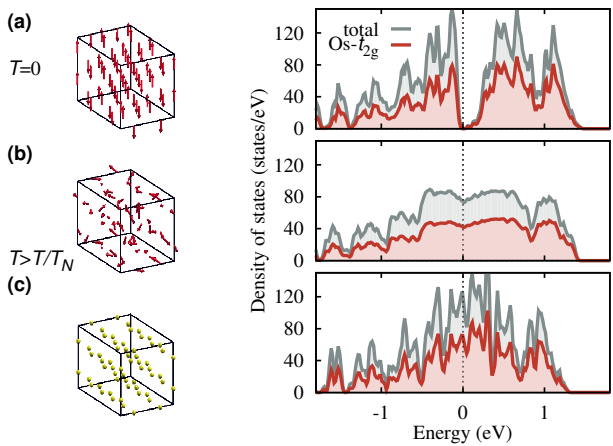}
\caption{(color online). {High-temperature paramagnetic state.}
Total and $t_{2g}$-projected  density of states (DOS)
of the ({a}) low-temperature insulating  AFM ground phase, ({b}) disordered paramagnetic phase (metallic)
and ({c}) nonmagnetic phase (metallic).
The arros indicate the Os sins.}
\label{para}
\end{figure}

\emph{Spin fluctuations and Lifshitz MIT.} After clarifying the actual correlation degree in NaOsO$_3$ and the
crucial role played by SOC to determine it,  we are ready to address the most
intriguing feature of this compound -- the origin of the continuous MIT  at
finite temperatures.

\begin{figure*}[t!]
\includegraphics[width=0.9\textwidth]{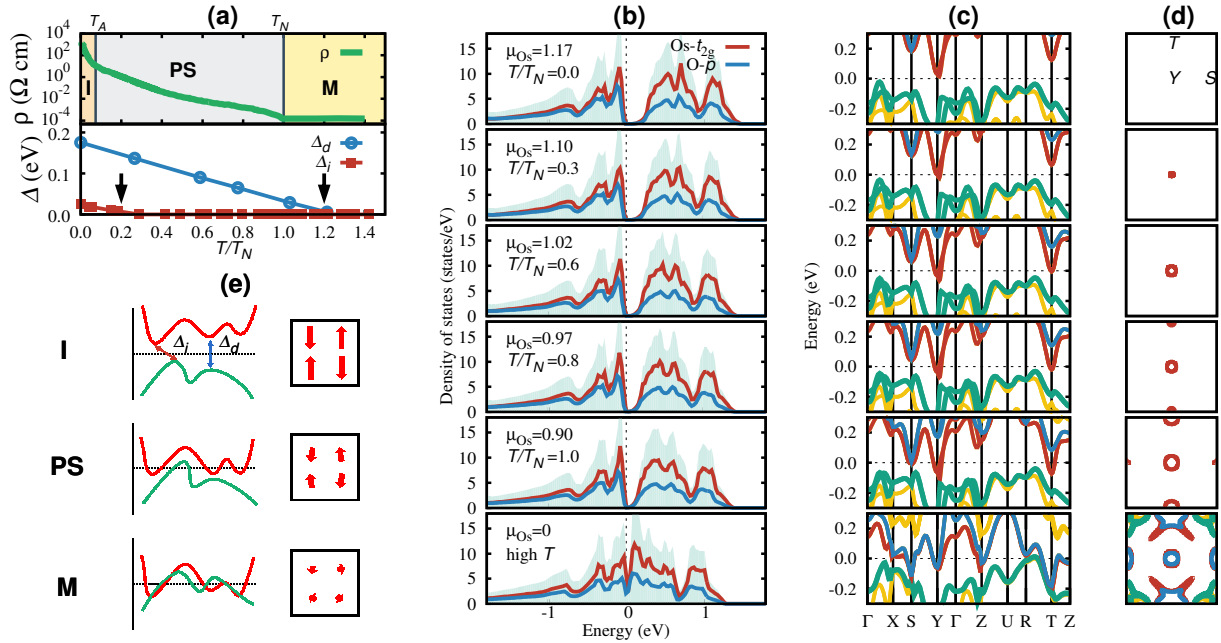}
\caption{(color online). {Temperature-dependent MIT.}
({a}) Indirect band gap  ($\Delta_i$) and direct pseudogap ($\Delta_d$)
as a function of temperature compared with the experimental resistivity curve readapted from
Ref. \cite{Shi2009}. The two anomalies in the resistivity curves at $T_A$ and $T_N$
set the transition from an insulating (I) to a pseudogap (PS) state and from the PS to a purely metal (M) state,
respectively. These two anomalies are correlated with the closing of the insulating ($\Delta_i$) and pseudo
($\Delta_d$) gaps (indicated by arrows).
({b}) Partial DOS ({c}) band structure and ({d}) FS for different temperatures ($T/T_N$)
corresponding to different Os magnetic moment $\protect\mu_{Os}$  (Eq. \ref{sft}).
({e}) Schematic diagram of the MIT. I: AFM insulator; PS: AFM pseudogap state with longitudinal (and small rotational)
spin fluctuation; M: magnetically itinerant metal.
}
\label{MIT}
\end{figure*}

Our main results are summarized in Figs.\ \ref{para} and
\ref{MIT}. First, we recall that the resistivity ($\rho$)
curve of NaOsO$_3$ (Fig. \ref{MIT}(a)) shows two anomalies: one at
$T_N=410$~K, corresponding to a sudden increase of $\rho$, and the second one
around $T_A$=30~K ($T/T_N \approx 0.1$)~\cite{Shi2009,Calder2012} characterized by a steeper increment.
The region below $T_A$ has a clear insulating-like behavior with a large and rapidly growing $\rho$,
whereas in the intermediate region, $T_A<T<T_N$, $\rho$ is always smaller then 1 ${\Omega}${cm} and
grows more slowly, with a bad-metal/pseudogap
behavior.
This hypothesis is supported by the experiments of Lo Vecchio \emph{et al.}
showing that the optical conductivity does neither vanish at
$T_N$~\cite{LoVecchio2013} nor shows a clear downturn at low
frequencies in the intermediate temperature region.
Such a peculiar temperature dependence of the conductivity is very unusual
for TMOs~\cite{LoVecchio2013}, where  -in the presence of a
Mott-Hubbard MIT- the opposite trend can be observed~\cite{Baldassarre2008}.

Similar temperature dependence properties were reported, instead, for the narrow-gap semiconductor FeSi:
at low temperature FeSi is a paramagnetic insulator but it develops a pseudogap associated with a bad-metal state
as temperature increases~\cite{Damascelli1997,Paschen1997,Mazurenko2010}. This anomalous behavior is explained well by spin fluctuations
in the context of the theory of itinerant magnetism elaborated by Moriya~\cite{Moriya1985}.
This similarity is of course very inspiring for the identification of the physical mechanisms at play in NaOsO$_3$.

To substantiate this idea we explore the effect of
transverse (rotational) and longitudinal spin fluctuations by means of non-collinear fixed spin moment calculations.
As a first step,  we have simulated the effect of rotational spin fluctuations in the high-temperature spin
disordered configuration by performing non-collinear magnetic calculations
on large supercells containing 32 Os sites starting from randomly rotated Os spin moments in a paramagnetic arrangement ({\it{i.e.}},
the total magnetic moment is zero), by fixing the magnitude of the magnetic moments to the ground state value.
The results are summarized in Fig.~\ref{para}.
The starting point is the density of states (DOS) of the reference
collinear AFM  insulating state at $T=0$, Fig.~\ref{para}(a).
The paramagnetic DOS, Fig.~\ref{para}(b), clearly shows that above $T_N$ the system is an ordinary metal independently on the
value of the local magnetic moment.
By allowing a full relaxation of the moments the non-collinear paramagnetic state converges to a unrealistic metallic non-spin-polarized
state (Fig.~\ref{para}(c)) with entirely quenched local spin moments, similarly to the case of other magnetically itinerant metals like Cr and Ni~\cite{Kubler}.
However, longitudinal spin fluctuations induce local moments forming the true realistic high-temperature paramagnetic state~\cite{Khmelevskyi2016}.
Thus, to explain the double anomalies behavior in the resistivity curve at finite temperature
(Fig.~\ref{MIT}(a)), the effect of both rotational and longitudinal spin fluctuations should be taken into the account.
To quantify the amplitude of the moments induced by longitudinal spin fluctuations at finite temperature we employ the phenomenological theory of
Moriya, as described below~\cite{Moriya1985}.

Within spin-fluctuation theory there is a universal relation between the magnitude of the local mean square
fluctuating magnetic moment at the transition temperature ($M = \sqrt{<M^2(T_{N})>}$)
and the ground state local moment ($M_0$) expressed by the Moriya formula $ <M^2(T_{N})>=3/5 M^2_{0} $ \cite{Moriya1985,Mohn2006}.
By combining the Moriya formula with the Mohn and Wohlfarth approximation~\cite{Mohn2006}, which assumes a linear temperature dependence of
the local mean square moment amplitude, we arrive to the following relation between the
amplitude of the fluctuating Os moment ($M(T)$) and temperature:
\begin{equation}
M(T)=M_{0}\sqrt{1-\frac{2}{5}\frac{T}{T_{N}}}.  \label{sft}
\end{equation}%
Using this relation we have conducted a series of calculations with fixed magnetic moments from $\mu_{Os} = M_0$ = 1.2 $\mu_B$ to 0 $\mu_B$ to examine
the changes of the electronic structure upon temperature.
The results are collected in Fig.~\ref{MIT}.  At $T=0$ (top panels) the system is an antiferromagnetic insulator with a narrow bandgap and an Os moment
$\mu_{Os}$=1.17$\mu_B$. As the temperature increases the resistivity
curve shows its first anomaly at  $T_A=T/T_N \approx$ 0.1
(Fig. \ref{MIT}(a)), corresponding to the closing of the indirect band gap $\Delta_i$ due to longitudinal spin fluctuations that pushes down
the conduction band minima at the $Y$ point (Fig.~\ref{MIT}(c)).
Right above $T_A$, the system starts to develop a FS (Fig. \ref{MIT}(d))
and enters a bad-metal state characterized by several hole and
electron pockets with a pseudogap (PS, see Fig.~\ref{MIT}(e)) separating
the valence and conduction bands. This pseudogap ($\Delta_d$) decreases with
temperature and finally closes for $M(T) =\mu_{Os} = 0.9\mu_B$ at
about $T \simeq T_N$, second anomaly in the resistivity curve indicated by an arrow in Fig.~\ref{MIT}(a).
This second anomaly, due to the rotational magnetic disorder described  previously, set the transition
to a metallic behavior for $T > T_N$. The entire process is sketched schematically in Fig. \ref{MIT}(e). \\

By refining our analyis, and considering the corresponding evolution
of the electronic bandstructure  (Fig.~\ref{MIT}(c)), we finally gain the complete
description of the MIT in NaOsO$_3$. The bands changes almost rigidly with temperature and lead to a continuous change of the FS
topology in terms of the appearance of progressively larger holes and electrons pockets (Fig.~\ref{MIT}(d)):
this represents a clear hallmark of a Lifshitz transition~\cite{Lifshitz1960, Sohn2015}.

Lifshitz transitions like the one described here are
likely to be relevant for other magnetic materials laying at the border
between a localized Mott picture and a metallic regime such as NiS~\cite{Panda2013}, or 5$d$ TMOs with a close to half-filled $t_{2g}^3$ configuration like
Cd$_2$Os$_2$O$_7$~\cite{Hiroi2015, Sohn2015} and iridates. Nearly half-filling, in fact, appears to be the optimum balance
between a localized (insulating) and deloclaized (metallic) scenario: a reduced filling is generally associated to strong $U$
picture~\cite{Liu2016} (i.e. $d^1$ Ba$_2$NaOsO$_6$ and  $d^2$ Ba$_2$CaOsO$_6$ are Mott-like insulator~\cite{Gangopadhyay}), whereas larger occupation increases the degree of
metallicity (i.e. $d^4$ BaOsO$_3$ and NaIrO$_3$~\cite{Jung2014}); moreover, the nearly $L_{eff}=0$ state reduces
the SOC-induced splitting which weaken the tendency towards Mott-SOC states and is functional for a rigid-band Lifshitz MIT.
Also, structural distortions in 5$d$ oxides are an important factor for Lifshitz MIT as they enables the formation of small local moments that
would be otherwise suppressed in undistorted and more strongly hybridized structures~\cite{Jung2013, Jung2014}.

\emph{Conclusions.} In summary, we have shown that the coexistence between weak electronic correlation and itinerant magnetism
can lead to a spin-driven Lifshitz MIT, {\it{i.e.}}, a magnetically induced continuous reconstruction of the Fermi surface. For NaOsO$_3$
we found that the MIT is prompted by transverse and longitudinal fluctuations of the itinerant Os moments, and is
intimately bound to the SOC-driven renormalization  of the electronic kinetic energy.
Our study explains the three different regimes observed experimentally: (i) At low-$T$ NaOsO$_3$ is an AFM insulator; (ii) At $T=T_A$
it enters a bad metal regime due to longitudinal spin fluctuations; finally (iii) at $T=T_N$ rotational spin fluctuations closes
the direct pseudogap and the system become a paramagnetic metal.
Spin fluctuations and the tunability by doping and pressure of the Lifshitz FS reconstruction could give rise to novel
magnetic, orbital and superconducting transitions, to be exploited for engeneering TMOs based heterostructures.
Moreover, the surprising role played by the SOC in NaOsO$_3$, as well as its ``competitive'' action
against correlations -highlighted here for the first time-
could radically affect the theoretical description of the relativistic
mechanisms controlling the electronic properties of many other oxides.

\section*{Acknowledgements}
We thank A. Perucchi, S. Lupi and L. Boeri for helpful discussions.
This work was supported by the joint FWF and Indian Department of Science and Technology (DST) project INDOX (I1490-N19), and
by the FWF-SFB ViCoM (Grant No. F41). Computing time at the Vienna Scientific Cluster is greatly acknowledged.

\pagebreak
\widetext
\begin{center}
\textbf{Supplementary information: A new route to relativistic insulators: Lifshitz transition driven by spin fluctuations and spin-orbit renormalization in NaOsO$_3$}
\end{center}

\renewcommand{\thefigure}{S\arabic{figure}}
\renewcommand{\thetable}{S\arabic{table}}
\renewcommand{\theequation}{S\arabic{equation}}
\renewcommand{\bibnumfmt}[1]{[S#1]}
\renewcommand{\citenumfont}[1]{S#1}

\setcounter{equation}{0}
\setcounter{figure}{0}
\setcounter{table}{0}
\setcounter{page}{1}
\makeatletter

\section{Computational Details}

\begin{figure}[b]
\includegraphics[width=0.45\textwidth]{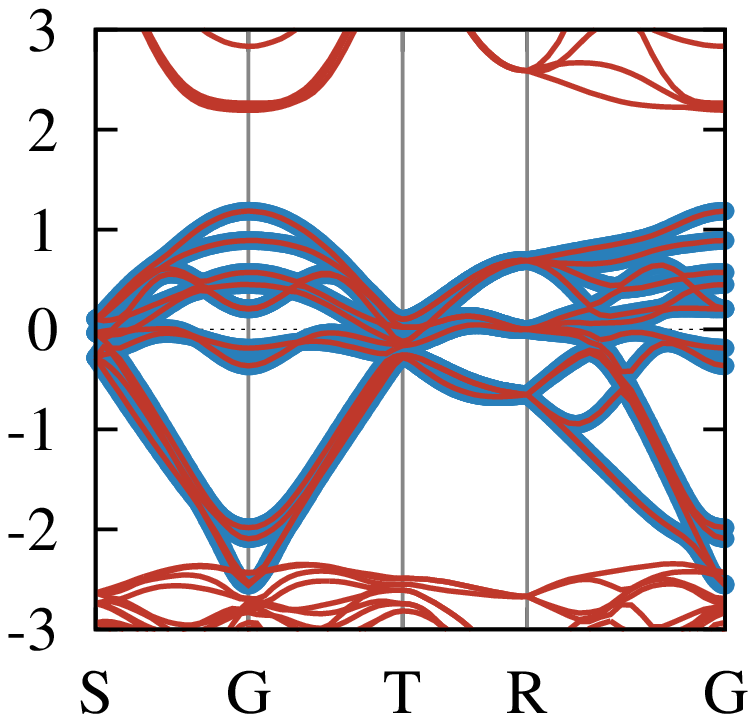}
\caption {Superimposition of the DFT and $t_{2g}$ wannier-projected band structure.
}
\label{fig:wannier}
\end{figure}

All calculations were performed using the Vienna \emph{ab initio} simulation package (VASP) \cite{Kresse1993,Kresse1996} using the
DFT+$U$ method~\cite{Dudarev1998} within the generalized gradient approximation and including spin-orbit interaction
(with quantization axis along the (0, 0, 1) direction).
The on-site Coulomb $U$ was computed fully {\em ab~initio} using the constrained random phase approximation (cRPA)~\cite{cRPA}.
For the cRPA calculations  we used a maximally-localized Wannier functions basis including the $t_{2g}$ states
that ensured an excellent match with the DFT band structure around the Fermi energy as shown in Fig.~\ref{fig:wannier}).
We used $U=0.4$~eV, slightly smaller than the value estimated by the cRPA, 0.68 eV (see main text).
We adopted the generalized gradient approximation, $U$=0.4~eV, plane-wave cutoff of 400 eV and a $t_{2g}$ basis set for cRPA.
The atomic position were fully relaxed at the experimental volume.
A 10$\times$6$\times$10 Monkhorst-Pack mesh was used (4$\times$3$\times$4 and 20$\times$20$\times$20
for non-collinear calculations and plasma frequency calculations, respectively).
The lattice parameters were fixed to the experimental ones and internal positions of all atoms were fully relaxed.
A 10$\times$6$\times$10 Monkhorst-Pack mesh was used (reduced to 4$\times$3$\times$4 for non-collinear magnetic calculations
in the large supercell containing 32 Os sites) with a plane-wave energy cutoff of 400 eV.
In order to estimate the role of relativistic effects on the electron correlation (relativistic renormalization) we
have followed the standard procedure described in Ref.\onlinecite{Qazilbash2009} which involves the calculation of
the plasma frequency in the paramagnetic phase using $U=0$. For these calculations  we increased the $k$-point grid up to
20$\times$20$\times$20 and very stringent energy convergence criteria ($10^{-8}$ eV).

To model the high-temperature paramagnetic phase we adopted a supercell containing 32 Os sites starting from randomly
distributed magnetic moments within a non-collinear set-up resulting in a zero total magnetic moment~\cite{Okatov2009,Wysocki2009,Glasbrenner2012}.
The degree of disorder was verified by computing the spin-spin correlation function S: for the random system we have obtain S=0.2, to be compared
with the ordered AFM system for which S=1.

\begin{thebibliography}{99}
\bibitem{Mott1968} N. F. Mott, Metal-Insulator
  Transition. \emph{Rev. Mod. Phys.} {\bf 40}, 677 (1968).
\bibitem{Imada1998}
 M. Imada, A. Fujimori and Y.  Tokura, Metal-insulator transitions, {Rev. Mod. Phys.} \textbf{70,} 1039 (1998).
\bibitem{Dagotto2005}
 E. Dagotto, Complexity in Strongly Correlated Electronic Systems, {Science} \textbf{309,} 257 (2005).
\bibitem{Tokura2000}
 Y. Tokura and N. Nagaosa, Orbital Physics in Transition-Metal
 Oxides, {Science} \textbf{288,} 462 (2000).
\bibitem{Kotliar2004} G. Kotliar and D. Vollhardt, Strongly Correlated
  Materials: Insights from Dynamical Mean-Field Theory, {Physics Today}
  {\bf 57}, 53 (2004).
\bibitem{Koster2012}
 G. Koster, L. Klein, W. Siemons, G. Rijnders, J. S. Dodge, C.-B. Eom, D. H. A. Blank, and M. R. Beasley,
 Structure, physical properties, and applications of SrRuO$_3$ thin films, {Rev. Mod. Phys.} \textbf{84} 253 (2012)
\bibitem{Cao1998}
 G. Cao, J. Bolivar, S. McCall, J. S. Crow, and R. P. Guertin,
 Weak ferromagnetism, metal-to-nonmetal transition, and negative differential resistivity in single-crystal Sr$_2$IrO$_4$, {Phys. Rev. B} \textbf{57,} R11039(R) (1998).
\bibitem{YKKim2015} Y. K. Kim, N. H. Sung, J. D. Denlinger, and
  B. J. Kim, Observation of a $d$-wave gap in electron-doped
  Sr$_2$IrO$_4$, Nature Physics \textbf{12}, 37 (2016).
\bibitem{Jackeli2009} G. Jackeli and G. Khaliullin, Mott
    Insulators in the Strong Spin-Orbit Coupling Limit: From
    Heisenberg to a Quantum Compass and Kitaev Models, {Phys. Rev. Lett.} \textbf{102}, 017205 (2009).
\bibitem{Pesin2010} D. Pesin and L. Balents, Mott physics and band
  topology in materials with strong spin-orbit interaction, {Nature
  Physics} \textbf{6}, 376 (2010).
\bibitem{BJKim2008}
 B. J. Kim {\em {et al.}}, Novel $J_{eff}=1/2$ Mott State Induced by
 Relativistic Spin-Orbit Coupling in Sr$_2$IrO$_4$, {Phys. Rev. Lett.} \textbf{101,} 076402 (2008).
\bibitem{Martins2011} C. Martins, M. Aichhorn, L. Vaugier, and Silke
  Biermann,  Reduced effective spin-orbital degeneracy and
  spin-orbital ordering in paramagnetic transition-metal oxides:
  Sr$_2$IrO$_4$ versus Sr$_2$RhO$_4$, {Phys. Rev. Lett.} {\bf
    107}, 266404 (2011).
  \bibitem{Shi2009} Y. G. Shi {\em {et al.}}, Continuous metal-insulator transition of the antiferromagnetic perovskite NaOsO$_{3}$, {Phys. Rev. B} \textbf{80,} 161104(R) (2009).
\bibitem{Calder2012}
 S. Calder {\em {et al.}},  Magnetically driven metal-insulator transition in NaOsO$_3$, {Phys. Rev. Lett.} \textbf{108,} 257209 (2012).
 \bibitem{LoVecchio2013}
 I. Lo Vecchio, A. Perucchi, P. Di Pietro, O. Limaj, U. Schade, Y. Sun, M. Arai, K. Yamaura, and S. Lupi,
 Infrared evidence of a Slater metal-insulator transition in NaOsO$_3$, {Sci. Rep.} \textbf{3}, 2990 (2013).
 \bibitem{Calder2015}
S. Calder {\em {et el.}}, , Enhanced spin-phonon-electronic coupling in a 5d oxide, Nat. Commun. \textbf{6}, 8916 (2015).
 \bibitem{Slater1951}
 J. C. Slater, Magnetic Effects and the Hartree-Fock Equation, {Phys. Rev.} \textbf{82}, 538 (1951).
    \bibitem{Hiroi2015}
Hiroshi Shinaoka, Takashi Miyake, and Shoji Ishibashi,
Noncollinear magnetism and spin-orbit coupling in 5d pyrochlore oxide $\rm Cd_2Os_2O_7$,
Hiroshi Shinaoka, Takashi Miyake, and Shoji Ishibashi
Phys. Rev. Lett. {\bf 108}, 247204(1-4) (2012);
 Z. Hiroi, J. Yamaura, T. Hirose, I. Nagashima, and Y. Okamoto,
 Lifshitz Metal-insulator transition induced by the all-in/all-out magnetic order in the pyrochlore oxide Cd$_2$Os$_2$O$_7$, {APL Materials} \textbf{3,} 041501 (2015).
\bibitem{Sohn2015}
C. H. Sohn {\em {et al.}},
Optical spectroscopic studies of the metal-insulator transition driven by all-in-all-out.
magnetic ordering in 5d Pyrochlore, Phys. Rev. Lett. \textbf{115}, 266402 (2015).
   \bibitem{Lifshitz1960}
 I. M. Lifshitz, Anomalies of electron characteristics of a metal in the high pressure region, {Sov. Phys. JETP} \textbf{11,} 1130 (1960).
\bibitem{Jung2013}
 M. C. Jung, Y. J. Song, K. W. Lee, and W. E. Pickett,
 Structural and correlation effects in the itinerant insulating antiferromagnetic perovskite NaOsO$_3$, {Phys. Rev. B} \textbf{87}, 115119 (2013).
 \bibitem{Middey2014}
 S. Middey, Saikat Debnath, Priya Mahadevan, and D. D. Sarma, NaOsO$_3$: A high Neel temperature 5d oxide, {Phys. Rev. B} \textbf{89},
 134416 (2014).
\bibitem{Du2012}
 Y. Du, X. Wan, L. Sheng, J. Dong, and S. Savrasov, Electronic structure and magnetic properties of NaOsO$_{3}$, {Phys. Rev. B} \textbf{85,} 174424 (2012).
\bibitem{Kresse1996}
 G. Kresse and J. Furthm\"{u}ller, Efficient iterative schemes for \emph{ab initio} total-energy calculations using a plane-wave basis set, {Phys. Rev. B} \textbf{54,} 11169 (1996).
 \bibitem{Stewart2011}
 M. K. Stewart {\em {et al.}} , Optical study of strained ultrathin films of
 strongly correlated LaNiO$_3$, {Phys. Rev. B} \textbf{83}
 075125 (2011).
\bibitem{Qazilbash2009}
 M. M. Qazilbash, J. J. Hamlin, R. E. Baumbach, L. Zhang, D. J. Singh, M. B.  Maple, and D. N. Basov,
 Electronic correlations in the iron pnictides, {Nat. Phys.} \textbf{5,} 647–650 (2009).
\bibitem{Aryasetiawan2004}
 F. Aryasetiawan, M. Imada, A. Georges, G. Kotliar, S. Biermann and A. I. Lichtenstein,
 Frequency-dependent local interactions and low-energy effective models from electronic structure calculations, {Phys. Rev. B} \textbf{70,} 195104 (2004).
\bibitem{Gangopadhyay}
Shruba Gangopadhyay and Warren E. Pickett,
Interplay between Spin-Orbit Coupling and Strong Correlation Effects:
Comparison of Three Osmate Double Perovskites: $\rm Ba_2AOsO_6$ (A=Na, Ca, Y),
Phys. Rev. B \textbf{93}, 155126 (2016).
\bibitem{Liu2015}
 P. Liu, S. Khmelevskyi, B. Kim, M. Marsman, D. Li, X.-Q. Chen, D. D. Sarma, G. Kresse and C. Franchini,
 Anisotropic magnetic couplings and structure-driven canted to collinear transitions in Sr$_{2}$IrO$_{4}$ by magnetically constrained noncollinear DFT, {Phys. Rev. B} \textbf{92}, 054428 (2015).
\bibitem{Millis2005}
 A. J. Millis, A. Zimmers, R. P. S. M. Lobo, N. Bontemps and C. C. Homes, Mott physics and the optical conductivity of electron-doped cuprates, {Phys. Rev. B} \textbf{72}, 224517 (2005).
\bibitem{Si2009}
 Q. Si, Iron pnictide superconductors: Electrons on the verge, {Nat. Phys.} \textbf{5,} 5 (2009).
\bibitem{Giovannetti2014}
G. Giovannetti and M. Capone, Dual nature of the ferroelectric and metallic state in LiOsO$_3$, {Phys. Rev. B} \textbf{90,} 195113 (2014).
\bibitem{Rondinelli2008}
 J. M. Rondinelli, N. M. Caffrey, S. Sanvito, and N. A. Spaldin,
 Electronic properties of bulk and thin film SrRuO$_3$: Search for the metal-insulator transition, {Phys. Rev. B} \textbf{78,} 155107 (2008).
\bibitem{Mohn2006}
 P. Mohn,  \emph{Magnetism in the Solid State.} (Springer, 2006).
\bibitem{Baldassarre2008} L. Baldassarre {\em {et al.}}, S. Quasiparticle evolution and pseudogap formation in V$_2$O$_3$: An infrared spectroscopy study
 {Phys. Rev. B} {\bf 77}, 113107 (2008).
\bibitem{Damascelli1997}
 A. Damascelli, K. Schulte, D. van der Marel, M. F\"{a}th, A. A. Menovsky,
 Optical phonons in the reflectivity spectrum of FeSi, {Physica B} \textbf{230,} 787 (1997).
\bibitem{Paschen1997}
 S. Paschen, E. Felder, M. A. Chernikov, L. Degiorgi, H. Schwer, H. R. Ott, D. P. Young, J. L. Sarrao, and Z. Fisk,
 Low-temperature transport, thermodynamic, and optical properties of FeSi, {Phys. Rev. B} \textbf{56,} 12916 (1997).
\bibitem{Mazurenko2010}
 V. V. Mazurenko, A. O. Shorikov, A. V. Lukoyanov, K. Kharlov, E. Gorelov, A. I. Lichtenstein and V. I. Anisimov,
 Metal-insulator transitions and magnetism in correlated band insulators: FeSi and Fe$_{1-x}$Co$_x$Si, {Phys. Rev. B} \textbf{81,} 125131 (2010).
 \bibitem{Moriya1985}
 T. Moriya, \emph{Spin Fluctuations in Itinerant Electron Magnetism} (Springer Series in Solid-State Sciences, 1985).
\bibitem{Kubler}
J. K\"ubler, \emph{Theory of Itinerant Electron Magnetism} (Oxford Science Publications, 2000).
\bibitem{Khmelevskyi2016}
S. Khmelevskyi, First-principles modeling of longitudinal spin fluctuations in itinerant electron antiferromagnets: High N\'{e}el temperature in the V3Al alloy
{Phys. Rev. B} \textbf{94},  024420 (2016).
 \bibitem{Jung2014}
Myung-Chul Jung and Kwan-Woo Lee,
Electronic structures, magnetism, and phonon spectra in the metallic cubic perovskite BaOsO$_3$,
Phys. Rev. B \textbf{90}, 045120 (2014).
 \bibitem{Liu2016}
 Peitao Liu, Michele Reticcioli, Bongjae Kim, Alessandra Continenza, Georg Kresse, D.D. Sarma, Xing-Qiu Chen, Cesare Franchini,
 arXiv:1606.09112 [cond-mat.str-el].
\bibitem{Panda2013}
 S. K. Panda, I. Dasgupta, E. \c{S}a\c{s}i\v{g}lu, S. Bl\"{u}gel, and D. D. Sarma,
 NiS - An unusual self-doped, nearly compensated antiferromagnetic metal, {Sci. Rep.} \textbf{3}, 2995 (2013).


\end{thebibliography}

\begin{thebibliography}{99}

 \bibitem{Kresse1993}
 G. Kresse and J. Hafner, {Ab initio} molecular dynamics for liquid metals, {Phys. Rev. B} \textbf{47,} 558 (1993).
\bibitem{Kresse1996}
 G. Kresse and J. Furthm\"{u}ller, Efficient iterative schemes for \emph{ab initio} total-energy calculations using a plane-wave basis set, {Phys. Rev. B} \textbf{54,} 11169 (1996).
\bibitem{Dudarev1998}
 S. L. Dudarev, G.A.  Botton, S. Y. Savrasov, C. J. Humphreys, and A. P. Sutton,
 Electron-energy-loss spectra and the structural stability of nickel oxide: An LSDA+U study, {Phys. Rev. B} \textbf{57,} 1505 (1998).
\bibitem{cRPA} F. Aryasetiawan, K. Karlsson, O. Jepsen, and U. Sch\"onberger,
Phys. Rev. B \textbf{74}, 125106 (2006).
\bibitem{Qazilbash2009}
 M. M. Qazilbash, J. J. Hamlin, R. E. Baumbach, L. Zhang, D. J. Singh, M. B.  Maple, and D. N. Basov,
 Electronic correlations in the iron pnictides, {Nat. Phys.} \textbf{5,} 647–650 (2009).
\bibitem{Okatov2009}
S. V. Okatov, A. R. Kuznetsov, Yu. N. Gornostyrev, V. N. Urtsev, and M. I. Katsnelson,
Effect of magnetic state on the $\gamma$-$\alpha$ transition in iron: First-principles calculations of the Bain transformation path,
Phys. Rev. B \textbf{79}, 094111 (2009).

\bibitem{Wysocki2009}
A.L. Wysocki, R.F.  Sabirianov, M. van Schilfgaarde, K.D. Belashchenko,
First-principles analysis of spin-disorder resistivity of Fe and Ni,
Phys. Rev. B \textbf{80}, 224423 (2009).

\bibitem{Glasbrenner2012}
J.K. Glasbrenner, K.D. Belashchenko, J. Kudrnovsky, V. Drchal, S. Khmelevskyi, I. Turek,
First-principles study of spin-disorder resistivity of heavy rare-earth metals: Gd-Tm series.
Phys. Rev. B \textbf{85}, 214405(2012).


\end{thebibliography}
\end{document}